\documentclass[letterpaper,12pt]{JHEP3}
\usepackage{amssymb}
\usepackage{mathrsfs}
\usepackage{cite}
\usepackage{texdraw}
\usepackage[english]{babel}
\usepackage{epsfig}

\newcommand{\be}{\begin{equation}}
\newcommand{\ee}{\end{equation}}
\newcommand{\bea}{\begin{eqnarray}}
\newcommand{\eea}{\end{eqnarray}}
\newcommand{\del}{\partial}
\newcommand{\f}{\frac}
\newcommand{\e}{\epsilon}

\newcommand{\g}{\gamma}

\newcommand{\hs}[1]{\hspace{#1 mm}}

\let\bm=\bibitem
\def\arsinh{\mathop{\rm arcsinh}\nolimits}
\def\fs{|\phi|^2}
\def\del{\partial}
\let\la=\label
\def\nn{\nonumber}

\newcommand{\w}[1]{\\[0.#1cm]}

\def\a{\alpha}
\def\b{\beta}
\def\c{\gamma}

\def\d{\delta}

\def\e{\epsilon}
\def\ve{\varepsilon}
\def\f{\phi}

\def\k{\kappa}
\def\l{\lambda}
\def\L{\Lambda}
\def\m{\mu}
\def\n{\nu}
\def\r{\rho}
\def\s{\sigma}

\def\t{\tau}
\def\th{\theta}

\def\x{\xi}
\def\X{\Xi}

\def\o{\omega}

\def\lra{\leftrightarrow}

\def\g{1+\e |\phi|^2}

\def\fs{|\phi|^2}
\def\ra{\rightarrow}
\def\lra{\leftrightarrow}

\title{ New Supersymmetric Solutions in $N=2$ Matter Coupled 
$AdS_3$ Supergravities}
\author{Nihat Sadik Deger \\ 
Department of Mathematics \\
Bogazici University \\
34342, Bebek, Istanbul, Turkey \\
\\
Feza G\"{u}rsey Institute \\
Emek Mah. No:68, Cengelkoy \\
34680, Istanbul, Turkey \\
E-mail: \email{sadik.deger@boun.edu.tr}}

\author{\"{O}zg\"{u}r Sar{\i}o\u{g}lu \\ 
Department of Physics \\
Middle East Technical University \\ 
06531, Ankara, Turkey \\
E-mail: \email{sarioglu@metu.edu.tr}}

\abstract
{We construct new 1/2 supersymmetric solutions in $D=3, N=2$, matter  
coupled, $U(1)$ gauged supergravities and study some of their properties.
We do this by employing a quite general supersymmetry breaking condition,
from which we also redrive some of the already known solutions. Among the 
new solutions, we have an explicit non-topological soliton for the 
non-compact sigma model, a locally flat solution for the compact 
sigma model and a string-like solution for both types of sigma 
models. The last one is smooth for the compact scalar manifold.} 

\keywords{Supergravity models, $AdS/CFT$ correspondence, Solitons}

\preprint{}

\begin{document}

\section{Introduction}

Supersymmetric solutions in supergravity theories have been quite 
fundamental
in understanding various aspects of string/M theory. However, such
solutions were not studied much for supergravities coupled to
non-linear sigma models due to their complexity. Because of its
relative simplicity, $D=3$ supergravities provide a good framework for 
understanding such systems. This has the further advantage from the
$AdS/CFT$ perspective \cite{mal1,mal2,mal3} since the dual theory
would be a two dimensional $CFT$.
 
With these motivations in mind, we find new supersymmetric solutions 
in the matter coupled $D=3, N=2$, $U(1)$ gauged supergravities and
study some of their properties. This model was constructed 
in \cite{ads2} and admits both compact and non-compact sigma model 
manifolds.  
There is also a well-defined flat sigma model limit. The theory contains 
only a Chern-Simons gauge field and no Maxwell term. The first 
supersymmetric solutions of this model were constructed in \cite{ads2}
and these described static, uncharged strings. Later the charged, 
stationary
generalizations of these strings superposed with gravitational and
Chern-Simons electromagnetic waves were obtained in \cite{string}.
Another class of solutions representing vortices were found in 
\cite{sam1},
where the model we consider was modified with a Fayet-Iliopoulos term. 
This changes the potential so that only topological solitons, by which 
we mean smooth solutions that interpolate between $AdS$ and Minkowski 
vacua, are allowed.

The supersymmetric vortices \cite{sam1} and strings \cite{string} were
obtained by using structure-wise similar supersymmetry breaking 
conditions.
In this paper, we consider a more general supersymmetry breaking condition 
which contains these previous cases and succeed in obtaining the Killing 
spinor
explicitly. In addition to the already known solutions, this also leads us 
to new ones. Among these, we have an explicit non-topological soliton 
solution (a smooth solution that approaches to $AdS$ vacuum when $|\f|=0$) 
for the non-compact sigma model, a locally flat solution for the compact 
sigma model and a string-like solution for both types of sigma models.
The last one is smooth for the compact scalar manifold.

The plan of this paper is as follows. In section 2 we begin with a review 
of the $N=2$ gauged supergravity model that we consider. In section 3 we 
give a
detailed analysis of the equations that arise from the supersymmetry
breaking condition. Section 4 is devoted to the construction of the new
supersymmetric solutions mentioned above. Solutions in the flat sigma
model are studied in section 5. We conclude in section 6 with some 
comments and future directions. The supersymmetry breaking condition is
worked out in appendix A.

\section{The Model}

In this paper we consider $N=2$, $U(1)$ gauged supergravity in $D=3$ 
interacting with an arbitrary number of matter multiplets which was 
constructed in \cite{ads2}. Its higher dimensional origin is yet to 
be discovered. The boundary symmetries of this theory were studied 
in \cite{mc2} and its extension by including a Fayet-Iliopoulos term 
was given in \cite{sam1}. Holographic RG flows in this model were 
analyzed in \cite{rg}. Let us also mention that the model we consider in 
this
paper \cite{ads2} is a member of a class of theories called abelian
Chern-Simons Higgs models coupled to gravity (see \cite{sam1, verbin} and
references therein). The field content of the theory is:
\

\, $\bullet$ The supergravity multiplet: \{$e_\m{}^a$, $ {\tilde \psi}_\m$, 
$A_\m$\}

\, $\bullet$ The scalar multiplet ($K$ copies): \{$\phi^\a$, 
${\tilde \l}^r$\}
\

\noindent
All fields except the graviton $e_\m{}^a$ and the gauge field $A_\m$ are 
complex. Here, for the sigma model manifolds we consider $K=1$ 
with the following cases, $M_{+}= S^2=SU(2)/U(1)$ and 
$M_{-}= H^2=SU(1,1)/U(1)$. We define the parameter $\e=\pm$1 to indicate 
the manifolds $M_\pm$. The bosonic part of the Lagrangian is
\footnote{Our conventions are as follows: We take $\eta_{ab} = (-,+,+)$ 
and 
$\e^{\m\n\r}=\sqrt{-g}\c^{\m\n\r}$. In a coordinate basis a convenient 
representation for $\gamma^{a}$ matrices is
$\gamma_0=i\sigma^3, \gamma_1=\sigma^1, \gamma_2=\sigma^2$ 
with $\epsilon^{012}=1$. Here 0,1,2 refer to the tangent time, radial and 
theta directions, respectively, and $\c^2$ is the charge conjugation 
matrix.} 
\be
{\cal L} = \sqrt{-g}\left(
{1\over 4} R
-{1\over 16ma^4}\, {\e^{\m\n\r} \over \sqrt{-g}} A_\m \del_\n A_\r 
- { |D_\m\f|^2 \over a^2(1+\e |\f|^2)^2} 
-V(\f) 
\right)\, ,
\la{ba1}
\ee
where $D_\m\f=(\del_\m -i\e A_\m)\f$ \, and
the potential is given by
\be
V(\f)= 4m^2a^2C^2\left( |S|^2-{1\over 2a^2}C^2\right)\, .
\la{a2}
\ee
Functions $C$ and $S$ are defined as 
\be
C= {1- \e\fs \over \g}\ , \hs{5} S = {2\f \over\g}\ .
\la{cs2}
\ee
\FIGURE{
\centerline{\epsfxsize=3.3truein
\epsffile{fig1.eps}
\hspace{0.4in}
\epsfxsize=3.3truein
\epsffile{pot2.eps}
}
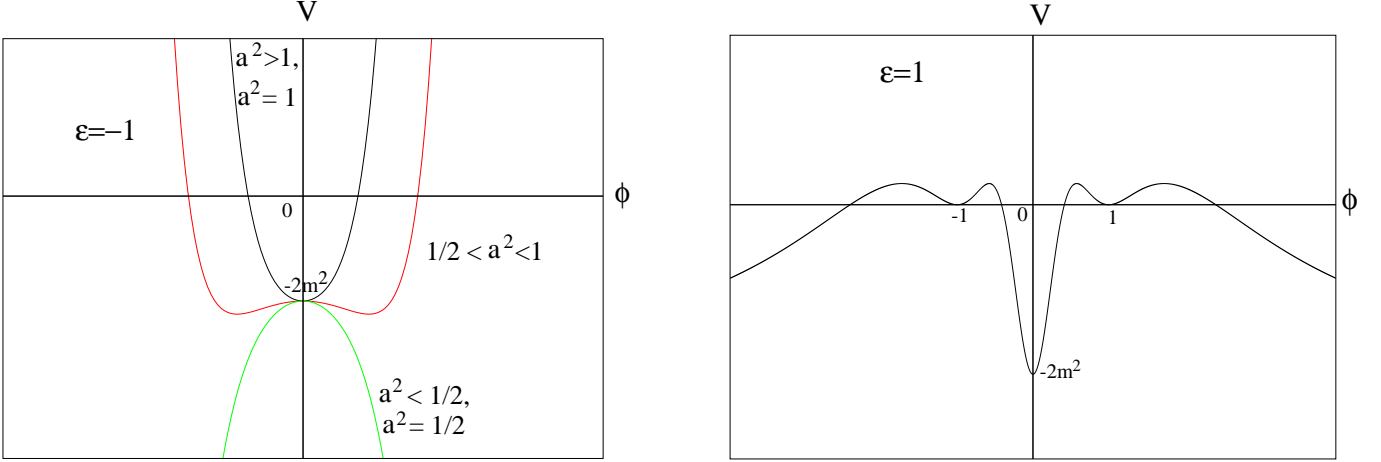
\caption{The scalar potential $V$ plotted with respect to $\phi$. For 
$\e=-1$, \, $V(|\f|=1) \to \infty$.}
}

Note that the following algebraic and differential relations hold:
\be
|\f|^2=\frac{\e (1-C)}{ (1+C)} \, , \hs{5} \e |S|^2=1-C^2 \, , 
\hs{5} \frac{dC}{d|\f|} = - \frac{\e |S|^2}{|\f|} \, , \hs{5} 
\frac{d|S|}{d |\f|} =  \frac{C|S|}{|\f|} \, .
\la{algrel}
\ee
The constant $``a"$ is the characteristic
curvature of $M_\pm$ (e.g. $2a$ is the inverse radius in the case
of $M_+=S^2$).
The gravitational coupling constant $\k$ has been
set equal to one and $-2m^2$ is the $AdS_3$
cosmological
constant.
Unlike in a typical $AdS$ supergravity coupled to matter,
the constants $\k,a,m$ are not related to each other for
non-compact scalar manifolds, while $a^2$ is quantized in terms of
$\k$ in the compact case so that $\k^2 /a^2$ is an
integer \cite{ads2}. When $\e=-1$ for all $a^2$ there is a supersymmetric 
$AdS$ vacuum at $\f=0$ and a non-supersymmetric but stable (it satisfies 
Breitenlohner-Freedman bound \cite{breit}) $AdS$ vacuum for $1/2<a^2<1$. 
When $\e=1$ there are supersymmetric $AdS$, Minkowski and  
non-supersymmetric de Sitter vacua (see Figure 1). The forms of the
potentials are appropriate for the possibility of existence of topological
($\e=1$) and non-topological ($\e=-1$) solitons.

The nonlinear scalar covariant derivative $P_\m$ and the $U(1)$ connection
$Q_\m$ are defined as
\be
P_\m = \frac{2 \del_\m\f}{\g}  -i\e  A_\m S \,, \quad
Q_\m = \frac{i \f \stackrel{\lra}{\del_\m}\f^*}{\g}\,+A_\m C\, .
\ee

The bosonic field equations that follow from the Lagrangian (\ref{ba1}) 
are
\bea
R_{\m\n} &=& \frac{1}{a^2} \,P_{(\m} P_{\n)}^*
+ 4 V g_{\m\n} \ ,
\label{e1}
\w2
\e^{\m\n\r}  F_{\n\r} &=&- 4\epsilon im a^2 \sqrt{-g}\,
[P^\m S^* - (P^{\m})^*S]\ ,
\label{e2}
\w2
\frac{1}{\sqrt{-g}} \del_\m \left(\sqrt{-g} g^{\m\n} P_\n\right)
&=& i\e Q_\m P^\m+ 2a^2 \left(1+\e |\f|^2\right)  \frac{\del 
V}{\del\phi^*}\, .
\label{e3}
\eea

The supersymmetric version of the Lagrangian (\ref{ba1}) is invariant 
under the following fermionic supersymmetry transformations 
\bea
\la{susy1}
\d {\tilde \psi}_\m &=& \left(\del_\m + \frac{1}{4} \o_\m{}^{ab} \c_{ab}
-\frac{i}{2 a^2}\,Q_\m \right) \ve +m \c_\m C^2 
\ve\ , \w2
\d{\tilde \l} &=&\left(- \frac{1}{2a} \c^\m P_\m - 2\e m a \, CS \right) 
\ve\, .
\la{susy2}
\eea
In order to obtain a supersymmetric solution, one needs to solve 
(\ref{susy1}) and (\ref{susy2}) when all fermions are set to zero.
It should be noted that the solutions can be naturally divided 
into two classes: $a^2\neq 0$ and $a^2=0$. (To take the $a^2=0$ limit, 
certain field rescalings should be performed as explained in section 5.) 
With these preliminaries
at hand, we now look for supersymmetric solutions of this model in 
the next section.

\section{Supersymmetry Analysis ($a^2 \neq 0$)}

Our metric ansatz is
\be
ds^2= -F^2dt^2 + H^2(Gdt+d\theta)^2+ dr^2 ,
\la{metric}
\ee
where $F,G$ and $H$ are functions of $r$ only. 
We choose the scalar field to be of the form
\be
\f=R(r)e^{in\theta}e^{i\l t}\, , 
\la{ans1}
\ee
where $n$ and $\l$ are real constants. For the vector field we pick the 
following gauge 
\be
A_\m=(A_t, A_r, A_\theta)=(\psi(r),0, \chi(r))\, .
\la{ans2}
\ee

For a supersymmetric solution, we look for Killing spinors that satisfy
$\d {\tilde \psi}_\m =0$ and $\d {\tilde \l} = 0$. For this purpose, we 
assume a
projection of the form \( (1-\c^a b_a)\ve= 0, \)
where $b_a$'s are some \emph{complex} functions that satisfy
\( b^a b_a=1 \). Since the analysis involves a long calculation, 
for purposes of readability, we will save those 
technical details to the appendix \ref{appen} and now carry on with the 
final outcomes of that investigation. After a careful study, one ends up 
with the following set of equations:
\bea
\la{eqn4}
(FHZ)' &=& 4mC^2 FH \, , \\
\la{eqn5}
G &=& g_0 + \frac{FZ}{H} \, , \\
\la{eqn1}
\l - \e\psi &=& \frac{1}{C}(\l -2\e a^2 c_2)+ qk\e a^2 \frac{H'}{CF} 
g_0 + 4qk\e ma^2C \, , \\
\la{eqn2}
n - \e\chi &=& \frac{1}{C}(n-2\e a^2 c_1) + qk\e a^2 \frac{H'}{CF} \, , \\
\la{eqn3}
\frac{R'}{R} &=& qk \frac{(n- \e \chi)}{FH} + 4 \e m a^2 CZ \, , \\
\la{eqn6}
\l - \e \psi &=& g_0 (n - \e \chi) + 4 qk \e ma^2 C \, .
\eea
Here $\l, k, n, g_0, c_1, c_2$ are arbitrary real constants 
and $q^2=1$. Throughout, we use prime to indicate differentiation with 
respect to $r$. Note that (\ref{eqn6}) relates the constants as 
\( \l - 2\e a^2c_2= g_0(n-2\e a^2c_1) \). Here we would like
to emphasize that for electromagnetic ``self-dual'' solutions where
$E=-g_0 B$, we need $g_0 \neq 0$ and $k=0$ as have already
been found in \cite{string}. The function $Z$ is defined as:
\be
Z \equiv \frac{p}{F}\sqrt{F^2-k^2} \, , \hs{5} p^2=1 \, .
\label{defnz}
\ee

Now one has to check the field equations. It turns out that the scalar 
field equation (\ref{e3}) is identically satisfied as a result of the
above set of equations. However, the vector field equation (\ref{e2})
yields one new equation 
\be
a^2 \left(\frac{H'}{F} \right)' + \frac{|S|^2 (n -\e 
\chi)^2}{FH} + 16m^2a^4C^2|S|^2\frac{H}{F} =0 \, .
\label{einstein}
\ee
On the other hand, after some rather lengthy but straightforward
calculations, it can be shown that the Einstein field equations (\ref{e1}) 
follow from (\ref{eqn4}), (\ref{eqn5}) and (\ref{einstein}).

In summary, for a 1/2 supersymmetric solution we need to solve equations 
(\ref{eqn3}), (\ref{eqn4}) and (\ref{einstein}) and determine the radial 
dependences of $H$, $F$ and $R$. (Note that $(n-\e\chi)$ can be replaced 
in these equations using (\ref{eqn2}).) Once this is done, the vector field 
components are determined. After some algebra, it can be shown that these 
three equations are equivalent to the following set:
\bea
\la{new1}
\frac{|S|'}{|S|}&=& \frac{\e C \sqrt{16m^2a^4C^2(FH)^2 + W}}{FH} \, , \\
\la{new2}
\frac{|S|'}{|S|}&=& \frac{\e a^2 (FH)'}{FH} +\frac{(n- 2\e a^2 
c_1)qk}{FH} \, , \\
\la{new3}
W' &=& 32\e m^2a^2(n- 2\e a^2 c_1)qk C^2 FH \, ,
\eea
where we defined
\be
\la{new4}
W \equiv k^2 (n-\e \chi)^2 + 8 \e ma^2qk CFHZ (n-\e \chi) - 
16m^2a^4k^2 C^2H^2 \, .
\ee

When $k=0$, instead of equations (\ref{new3}) and (\ref{new4}), we have 
(\ref{einstein}). Comparing (\ref{new3}) and (\ref{eqn4}), we see that
\be
W=8\e m a^2(n- 2\e a^2 c_1)qk FHZ + w_0 \, ,
\la{new5}
\ee
where $w_0$ is a constant that vanishes when $k=0$. Equations (\ref{new4}) 
and (\ref{new5}) together give an implicit relationship between the 
unknown functions $\chi, R, F$ and $H$. It is interesting to observe 
that the function $W$ can be related to a topological invariant as follows:
\be
F_{\m\n} F^{\m\n} = - \frac{32 m^2 a^4 |S|^4 W}{F^2 H^2} \, .
\la{chern}
\ee
Note that there is a crucial difference between the $k=0$ and $ k \neq 0$ 
cases in terms of the above invariant; for $k=0$, it vanishes automatically.

With these, our metric becomes:
\be
ds^2= -k^2dt^2 + 2p FHZ dvdt + H^2dv^2 + dr^2 \, ,
\hs{5} v \equiv \th + g_0t \, .
\la{metricfinal}
\ee
The curvature scalar of this metric is:
\be
g^{\m\n}R_{\m\n}=-2\left( \frac{(FH)''}{FH} - 4m^2C^4 \right) \, .
\la{invariant}
\ee

The set of equations (\ref{new1}), (\ref{new2}), (\ref{new3})
is quite difficult to analyze in its most general form.
This system was also obtained in \cite{sam1} with $g_0= \l =
c_2=0$ and $k=1$, however with the potential modified with the
Fayet-Iliopoulos term. Note that the system simplifies when $k=0$. In this
case we have $Z^2=1$ and $W=0$, and the equations are completely
integrable. This solution was obtained in \cite{string} and it
corresponds to a charged, stationary string with gravitational and
Chern-Simons waves attached to it \cite{string}. Therefore, we will
assume $ k \neq 0$ in this paper. There is another option
which simplifies this system, namely $(n- 2\e a^2 c_1)=0$ case.
Finally, when $R$ is a constant, that is, the scalar field is just a
phase, again the system is completely solvable. Now we analyze these, as
well as the general case in detail.

\section{Supersymmetric Solutions ($a^2 \neq 0$)}

In this section we will try to solve the set of equations 
(\ref{new1})-(\ref{new3}). We start with some easier subcases and 
consider the most general case later.

\subsection{$(n- 2\e a^2 c_1)=0$ Case}

In this case, we see from (\ref{new5}) that 
\be
W=w_0 \,,
\ee
where $w_0$ is an arbitrary real constant. Then we find from (\ref{new2})
that
\be
FH= f_0 |S|^{\e/a^2} \, ,
\la{fh}
\ee
which makes (\ref{new1}) a separable first order differential equation. 
Now we proceed with an investigation of the $w_0=0$ and $w_0 \neq 0$ 
cases separately.

\subsubsection{$w_0=0$}

When $w_0=0$, (\ref{new1}) is easily integrated \cite{string}, but its
explicit form will not be necessary for the discussion below. From 
(\ref{eqn4}), one finds
\be
Z= 1- \frac{u_0}{FH} \, ,
\label{z}
\ee
where $u_0$ is a real constant. Using (\ref{z}) and (\ref{fh}), one 
obtains
\be
H^2= \frac{f_0^2}{k^2} (2u_0 f_0 |S|^{\e/a^2} - u_0^2) \, .
\ee
For the vector field (\ref{eqn2}), one gets
\be
n - \e \chi = \frac{4 \e m a^2 u_0 f_0}{qk} C \, , \la{ppvec}
\ee
which is smooth for $\e=1$. The metric (\ref{metricfinal}) now becomes 
\be
ds^2= - k^2 du^2 + 2f_0 |S|^{\e/a^2} dudv + dr^2 \, , \hs{5} u 
\equiv pt+ \frac{u_0f_0}{k^2}v \, , \la{ppmet}
\ee 
which reduces to the string solutions (with no waves attached) 
that were found in \cite{string} when $k=0$. The scalar field in 
\cite{string} and the solution presented here is the same; however, 
$(n - \e \chi)$ in that case was proportional to $1/C$ from (\ref{eqn2}). 
This change makes the vector field smooth for $\e=1$. Unfortunately, we 
couldn't identify what this solution represents. However, because of the 
fact that this and the string solution presented in \cite{string} both 
have the same curvature invariants and the same scalar field, we call this 
as a `string-like' solution. Note that, for both solutions the $F^2$ 
invariant vanishes (\ref{chern}). The main difference of this new one is 
the absence of the $SO(1,1)$ worldsheet symmetry of the string solution 
\cite{string}. 

There is a curvature singularity (\ref{invariant}) as 
$C^2 \rightarrow \infty$ which can be seen from the curvature invariants
\bea
g^{\m\n}R_{\m\n} &=& -8m^2C^2(3C^2-8a^2|S|^2) \, , \nn \\
R^{\m\n}R_{\m\n} &=& 64 m^4 C^4(3C^4 - 16 a^2C^2|S|^2 + 24 a^4|S|^4) \, .
\la{ppinv}
\eea

When $\e=-1$, by performing a coordinate transformation \( \r=1/|S|^2
\;\; (0 \le \r < \infty) \), the metric (\ref{ppmet}) becomes 
\be
ds^2= - k^2 du^2 + 2f_0 \r^{1/(2a^2)} dudv + \frac{d \r^2}{64 m^2 a^4 
(\r+1)^2} 
\,,  \quad \e=-1 \, .
\ee
By inspection, it is seen that there is no horizon and we have a naked 
singularity at $\r=0 $ (or $C^2 \to \infty$). As $\r \to \infty$ (or $C^2 
\to 1$) the solution becomes locally $AdS_3$ whose metric corresponds to 
a generalized Kaigorodov metric \cite{kaig}.

Let us now consider the $\e=1$ case in more detail. We first define a new
radial coordinate \( \r=M/C^2 \;\; (M \le \r < \infty), \) where $M$ is 
a positive constant. Then (\ref{ppmet}) becomes
\be
ds^2= - k^2 du^2 + 2f_0 \left( 1- \frac{M}{\r} \right)^{1/(2a^2)} dudv + 
\frac{d \r^2}{64 m^2 a^4 (\r-M)^2} \,,  \quad \e=1 \, . \la{ppmet2}
\ee
We see that there is a horizon as $\r \to M$; from the curvature 
invariants
(\ref{ppinv}) the local geometry is observed to be locally $AdS_3$, which 
has 
a Kaigorodov \cite{kaig} type of structure. As $\r \to \infty$, the 
solution
is asymptotically flat. Since 
\( |\f|^2 = (\sqrt{\r}-\sqrt{M})/(\sqrt{\r}+\sqrt{M}) \), the scalar field
and the vector field (\ref{ppvec}) are smooth everywhere. 
Both fields have asymptotic values that are expected from a topological 
soliton, however the presence of a horizon prevents us from labeling this 
solution as such. 

Now let us look at the behavior of the geodesics. The geodesic 
equation associated with the metric (\ref{ppmet2}) is:
\be
\frac{1}{64m^2a^4}\left(\frac{\dot{\r}}{\r}\right)^2=
\a \left(1-\frac{M}{\r}\right)^2
-\frac{EP}{2 f_0} \left(1-\frac{M}{\r}\right)^{2-\frac{1}{2a^2}} - 
\frac{P^2 k^2}{4 f_0^2} \left(1-\frac{M}{\r}\right)^{2-\frac{1}{a^2}}  \, ,
\la{geodesic}
\ee
where the dot denotes derivative with respect to an affine parameter and
$\a=0$ or $\a=-1$ for null or timelike geodesics, respectively. In this
equation $E$ and $P$ are the conserved quantities associated with the flow
of the tangent vector of a geodesic corresponding to the $t$ and $v$
variables. It is easy to see that timelike geodesics can not reach 
the horizon since there is always a turning point. For null geodesics,
if one demands the geodesics to reach the $\r \ra \infty$ limit, one 
requires $-k^2 P^2 -2 EP f_0 > 0$. One can show that in this case there
is no turning point and such geodesics reach the horizon. However, for
$1/a^2=1,3 \; (\mbox{mod} \, 4)$, these geodesics can not extend beyond 
the horizon since (\ref{geodesic}) becomes imaginary. Moreover, for 
$1/a^2=2 \; (\mbox{mod} \, 4)$, it is easy to see that the geodesics 
do not cross the horizon since the right hand side of (\ref{geodesic}) 
becomes negative then. Therefore, the solution is well-defined for all
values of $1/a^2$ except when $1/a^2 = 0 \; (\mbox{mod} \, 4)$.

\subsubsection{$w_0 \neq 0$ and $\e=-1$}

For general $a^2$, when $w_0 \neq 0$, the integration of (\ref{new1}) is 
quite complicated. Therefore we will mainly concentrate on the $a^2=1$ case. 
In this case, we first introduce a new constant $|b| \le 1$ such that
$w_0 = - 16 m^2 b^2$, to simplify the discussion. Now (\ref{new1})
implies that
\[ \frac{R'}{R} = 4 m \sqrt{C^2 - b^2 |S|^2} \, . \]
From this it follows that
\[ dr = \frac{dC}{4m |S|^2 \sqrt{C^2 - b^2 |S|^2}} \, . \]
Furthermore, introducing $U \equiv FHZ$ and taking $U=U(C)$, one finds
from (\ref{eqn4}) that
\[ \frac{dU}{dC} = f_0 \frac{C^2}{(C^2-1)^{3/2} 
\sqrt{C^2 - b^2 (C^2-1)}} \,, \]
which can be integrated using an elliptic function $E$ \footnote{The 
elliptic function of the 2nd kind is defined as 
\( E(\f|m) = \int_{0}^{\f} (1-m \sin^2{\theta})^{1/2} d\theta ,\)
$\f \in (-\pi/2,\pi/2)$.} as
\[ U(C) = b f_0 \frac{-C \sqrt{1+(-1+\frac{1}{b^2})C^2} + \sqrt{1-C^2} 
E(\arcsin{C}|1-\frac{1}{b^2})}{\sqrt{C^2-1}} \, . \]
At first sight, this seems to be complex valued, however it can be verified 
that $U(C)$ is always real for $C>1$ (which is automatically satisfied for 
$\e=-1$) and $|b|<1$. With these, the metric can be cast in the form
\[ ds^2 = -(pk dt - U(C) dv)^2 + f_0^2 \frac{dv^2}{|S|^2} +
\frac{dC^2}{16 m^2 |S|^4 (C^2 - b^2 |S|^2)} \, . \]
Note that $C^2 - b^2 |S|^2 = 0$ when $C^2=b^2/(b^2-1)<1$, which is
not allowed and therefore there is no horizon! The curvature invariants 
for this metric are
\begin{eqnarray*}
g^{\m\n} R_{\m\n} & = & -8 m^2 (8 C^2 -5 C^4 + 4 b^2 |S|^4) \,, \\
R^{\m\n} R_{\m\n} & = & 64 m^4 [8 b^4 |S|^8 - 8 b^2 C^2 |S|^4 (2 |S|^2 -1)
+ C^4 (24 -32 C^2 + 11 C^4)] \,,
\end{eqnarray*}
from which it follows that as $C \to 1 (|S| \to 0)$, the solution becomes
locally $AdS$. The only place where a curvature singularity may appear is
at $C \to \infty$ (and thus $|S| \to \infty$). However, this may not always 
be an allowed limit. To see this, let us consider the $b=1$ case which
simplifies the calculations above. Then (\ref{new1}) becomes 
\( R'/R = - 4m \, .\) Defining a new radial coordinate
\( \r= 1/|S|^2 \,, \) we obtain
\be
FHZ= f_0 \sqrt{\r + 1} - \frac{f_0}{2} \ln \left( \frac{\sqrt{\r + 1} 
+1}{\sqrt{\r + 1} - 1} \right) \, .
\ee
From (\ref{fh}), we have $FH=f_0 \sqrt{\r}$ \, which yields,
\be
Z= \sqrt{\frac{\r + 1}{\r}}- \frac{1}{2 \sqrt{\r}} 
\ln \left( \frac{\sqrt{\r + 1} +1}{\sqrt{\r + 1} - 1} \right) \,.
\ee
Since we have $|Z| \leq 1$ from (\ref{defnz}), we see that this 
forbids $\r$ to reach 0. 

In terms of the new radial coordinate, $R=-\sqrt{\r}+\sqrt{1+\r}$,
which clearly shows that the scalar field is always smooth and finite. 
From (\ref{eqn3}), we obtain
\be
\chi= - n + \frac{4mf_0}{qk} \left[ \frac{1}{\sqrt{\r}} - 
\sqrt{\frac{\r+1}{\r}} 
\ln \left( \frac{\sqrt{\r + 1} +1}{\sqrt{\r + 1} - 1} \right) \right] \, ,
\ee
which is again smooth everywhere. This also implies the smoothness of 
$\psi$ from (\ref{eqn6}). All these matter fields have the expected
behavior of a non-topological soliton as $\r \to \infty$.

\subsubsection{$w_0 \neq 0$ and $\e=1$}

When $a^2=1$, the discussion is analogous to the one for the $\e =-1$ 
case: 
We start by introducing a new constant $b>0$ such that $w_0 = 16 m^2 b^2$. 
In this case (\ref{new1}) yields
\[ \frac{R'}{R} = 4 m \sqrt{C^2 + \frac{b^2}{|S|^2}} \, . \]
We record here in passing that now
\[ dr = \frac{dC}{4 m |S|^2 \sqrt{C^2 +\frac{b^2}{|S|^2}}} \, . \]
Introducing $U \equiv FHZ$ as before and taking $U=U(C)$ again, one 
obtains
from (\ref{eqn4}) that
\[ \frac{dU}{dC} = f_0 \frac{C^2}{\sqrt{C^2(1-C^2)+b^2}} \,, \]
whose integration yields the following complicated expression in terms of
elliptic functions $E$ and $F$ \footnote{The 
elliptic function of the 1st kind is given by 
\( F(\f|m) = \int_{0}^{\f} (1-m \sin^2{\theta})^{-1/2} d\theta ,\)
$\f \in (-\pi/2,\pi/2)$.}:
\begin{eqnarray*}
U(C) & = & \frac{i f_0 (\x+1) \sqrt{\x-1}}{2 \sqrt{2(C^2(1-C^2)+b^2)}} 
\sqrt{1+\frac{2C^2}{\x-1}} \sqrt{1-\frac{2C^2}{\x+1}} \times \\ 
& & \left(
E \left( i \arsinh{\left[ \sqrt{ \frac{2}{\x-1}} C \right]} \left|  
\frac{1-\x}{1+\x} \right. \right) -
F \left( i \arsinh{\left[ \sqrt{ \frac{2}{\x-1}} C \right]} \left| 
\frac{1-\x}{1+\x} \right. \right) \right) \, , 
\end{eqnarray*}
where we have used $\x \equiv \sqrt{1+4 b^2}>1$ for convenience. This 
may again seem to be complex valued, however it can be verified that 
$U(C)$ is always real for $0<C<1$ (which is automatically satisfied for 
$\e=1$). The metric now reads
\[ ds^2 = -(pk dt - U(C) dv)^2 + f_0^2 |S|^2 dv^2 +
\frac{dC^2}{16 m^2 |S|^2 (C^2 |S|^2 + b^2)} \, . \]
It is clear that there is no horizon. The curvature invariants for this 
metric are
\begin{eqnarray*}
g^{\m\n} R_{\m\n} & = & 8 m^2 (4 b^2 - 11 C^4 + 8 C^2) \,, \\
R^{\m\n} R_{\m\n} & = & 64 m^4 (8 b^4 - 8 b^2 C^2 (4 C^2 -3)
+ C^4 (24 - 64 C^2 + 43 C^4)) \,,
\end{eqnarray*}
which are both regular for $0<C<1$.

Looking at the curvature scalar above, one notices that this solution
approaches neither to $AdS$ ($C=1$ limit, where $g^{\m\n} R_{\m\n} = 
-24m^2$) nor the Minkowski vacuum ($C=0$ limit). Indeed, if one plots 
$(1-Z^2)=k^2/F^2$, one finds that this becomes negative before the 
$C=1$ point is reached. We thus conclude that this solution must be ruled 
out. 

\subsection{$R'=0$ Case}

For the sake of completeness, we also consider this case.
Looking at the supersymmetric vacua in Figure 1, we see that one should 
set $R=0$ when 
$\e=-1$, and $R=0$ or $R=1$ when $\e=1$. The scalar field is just a pure 
phase now. The functions $C$ and $|S|$ are just constants and by defining 
\[ \a \equiv - \frac{(n-2\e a^2 c_1)qk\e}{a^2} \quad \mbox{and} \quad
x_0 \equiv -4 m \left( \frac{1-\e R_0^2}{1+ \e R_0^2} \right)^2 \, ,\]
where $|\f|=R_0$, one finds from (\ref{new2}) and (\ref{eqn4}) that
\[ FH = \a r + \b \quad \mbox{and} \quad FHZ \equiv 
U(r) = - x_0 (\a \frac{r^2}{2} + \b r +u_0) \, , \]
for some integration constants $\b$ and $u_0$. The metric can now be 
cast in the form
\be 
ds^2 = -(pk dt - U(r) dv)^2 + (\a r + \b)^2 dv^2 + dr^2 \, , \la{metek}
\ee
whose curvature invariants simply read
\[ g^{\m\n} R_{\m\n} = \frac{x_0^2}{2} \quad \mbox{and} \quad 
R^{\m\n} R_{\m\n} = \frac{3}{4} x_0^2 \, , \]
which indicate that the metric describes a local $dS$ spacetime, when
$x_0 \neq 0$. This enforces us to set $x_0=0$, i.e. $R_0=1$, which is
only allowed for $\e=1$. This also sets $Z=0$, since now $U(r)=0$.
These imply from (\ref{eqn2}) that \( \chi = \e n .\) The metric 
(\ref{metek}) becomes (after absorbing constants in the 
metric by redefining the $t$ and the $v$ coordinates)
\be
ds^2= -dt^2 + r^2 dv^2 + dr^2 \, ,
\ee
which is a locally flat spacetime with Euclidean Rindler spatial sections. 

\subsection{General Case}

To analyze the general case (i.e. $(n-2\e a^2c_1) \neq 0$ and $k \neq 0$),
we first note that from equations (\ref{eqn4})-(\ref{eqn6}), one also 
finds
\be
\left[ FH \frac{R'}{R} \right]'= 16 \e m^2a^2 CFH (C^2-a^2 |S|^2) \, .
\la{newR}
\ee
Defining a new function $Y$ through the relation
\be
FH= f_0 |S|^{\e/a^2} Y \, ,
\ee
one can also show that
\be
\frac{Y'}{Y}= -\frac{qk\e(n-2\e a^2c_1)}{a^2 FH} \, .
\la{Y}
\ee
Using (\ref{Y}), we can write (\ref{newR}) as
\be
\frac{1}{Y} \frac{\del}{\del Y} \left[ Y \frac{\del \ln{R}}{\del Y} 
\right] 
= - \frac{\del V_{\rm eff}}{\del (\ln R)} =
\frac{16 \e m^2a^6 f_0^2}{(n-2\e a^2c_1)^2 k^2} 
C|S|^{2\e/a^2}(C^2-a^2 |S|^2) \, , \la{ydenk}
\ee
which decouples $R$ from other unknown metric functions as was observed
in \cite{sam1}. A simple integration yields the effective potential to be
\be
V_{\rm eff} = - \frac{8 m^2 a^8 f_0^2}{(n-2\e a^2c_1)^2 k^2 (\e+a^2)}
C |S|^{2\e/a^2} [\e C^2 + a^2 (C^2 - \e |S|^2)] \, .
\la{efpot}
\ee
These can be interpreted as describing a classical mechanical system where 
a (fictitious) point particle is subject to a motion due to an effective 
potential. 

When $\e=1$, it was shown in \cite{sam1} that there is no vortex solution 
where the scalar field approaches to a Minkowski vacuum as $r \to \infty$ 
and to an $AdS$ vacuum as $r \to 0$. To see this in our setup, we impose 
the following $AdS$ behavior around 
$r=0$: \( H \approx r \;, F \approx k . \) Then the regularity of the scalar 
field implies that $q=1, c_1=1/2$, and we get $\chi=0$ and $R=R_0 r^n$,
where $R_0$ is a constant. At the other end, as $r \to \infty$ demanding 
$R \approx 1$, $H \approx H_{\infty} r$, $FHZ = $ const., the regularity
of the vector field $\chi$ (\ref{eqn2}) imposes that 
$H_{\infty} = 1- n/a^2$. A well defined conical geometry requires 
$n/a^2 < 1$. However, this is not possible due to the fact that $1/a^2$ is 
an integer for $\e=1$. (When $n=0$, one is forced to set $|\f| =1$ 
\cite{cckk} which is analyzed in section 4.2.)

For $\e=-1$, there may be a non-topological soliton, however (\ref{ydenk}) 
is hard to analyze. In principle, one has to study (\ref{ydenk}) in 
the limits $C \to 1$ and $C \to \infty$ separately and match the two
solutions in a unique fashion. However, the limit $C \to 1$ 
(i.e. $R \to 0$) is quite difficult to work with due to the divergent 
right hand side (the exponent of $|S|$ is negative), however, techniques 
developed in \cite{cckk} might be applicable. If such a solution exists, 
then $C \to \infty$ shouldn't be accessible since this would make the vector 
field component $\psi$ divergent (\ref{eqn6}).

\section{Flat Sigma Model ($a^2=0$)}

To take the $a^2=0$ limit in our model, first one has to rescale
$A_\m \ra a^2 A_\m$ and $\f \ra a\f$. Then we have $C\ra 1$, $S \ra 2a 
\f$, and one obtains $N=2$, $AdS_3$ supergravity with cosmological 
constant $-2m^2$ coupled to an $R^{2}$ sigma manifold \cite{ads2}. 
This coincides with the flat sigma model limit of the $N=2$ theory 
discussed in \cite{it} as was shown in \cite{mc2}. The Lagrangian 
(\ref{ba1}) now becomes
\be
{\cal L} = \sqrt{-g}\left( \frac{1}{4} R
-\frac{1}{16m}\, \frac{\e^{\m\n\r}}{\sqrt{-g}} A_\m \del_\n A_\r 
- |\del_\m\f|^2 +2m^2 \right)\, ,
\ee
and its fermionic supersymmetry transformations are
\bea
\d {\tilde \psi}_\m &=& \left(\del_\m + \frac{1}{4} \o_\m{}^{ab} \c_{ab}
-\frac{i}{2}\,[ i \f \stackrel{\lra}{\del_\m}\f^* + A_\m] \right) 
\ve +m \c_\m 
\ve\ , \w2
\d{\tilde \l} &=& - ( \c^\m \del_\m \f )  \ve\, .
\eea

To find a 1/2 supersymmetric solution, we again choose the same metric 
ansatz (\ref{metric}) and use the same form of scalar and vector fields 
given in (\ref{ans1}) and (\ref{ans2}). Now using the projection condition 
(\ref{proj}) in $\d {\tilde \l} =0$ and $\d {\tilde \psi}_\m =0$, we find
\bea
R&=& R_0 = \mbox{const.} \, , \\ 
G &=& \frac{\l}{n} + \frac{FZ}{H}  \, , \\
\chi &=& 2c_1 - 2n R_0^2  \pm \frac{kH'}{F}  \, , \\
\psi &=& 2c_2 - 2\l R_0^2 \pm 4km \pm \frac{kH'}{F} \frac{\l}{n} \, .
\eea

The remaining unknown functions $F$ and $H$ are to be determined from
\bea
\la{a3}
(FHZ)' &=& 4mFH \, , \\
\la{a4}
\left(\frac{H'}{F}\right)' &=& -\frac{4R_0^2n^2}{FH} \, .
\eea
Here the first equation (\ref{a3}) comes from the supersymmetry analysis 
and 
the second one (\ref{a4}) follows from the field equations. When $k=0$, 
we have $Z^2=1$ (\ref{defnz}) and (\ref{a4}) is easily integrable then. 
This case was studied in \cite{string}. The metric in this case takes the 
form
\be
ds^2= -2pf_0e^{-4pmr}dvdt + (h_0 e^{-4pmr} + h_1 + h_2 r ) dv^2 + dr^2 \, 
, 
\label{meta=0}
\la{firstmetric}
\ee
where $f_0$ is an integration constant. Here $h_0$ and 
$h_1=(pmc_0+2R_0^2n^2)/4m^2$ (or $c_0$) are arbitrary real 
constants and we have \( h_2=-2R_0^2n^2p/m .\)

A study of the curvature invariants indicate that the solution has 
constant 
negative curvature $-24m^2$ and it is locally $AdS_3$ \cite{string}. 
When $h_1=h_2=0$, the metric is the $AdS_3$ metric in Poincar\'e 
coordinates. The $h_1$ term can be obtained by using the 
Garfinkle-Vachaspati method \cite{vac, gar} and it describes a wave 
in $AdS_3$. Actually the metric with $h_2=0$ has
already been discussed in \cite{pope} and it corresponds to a generalized
Kaigorodov metric \cite{kaig}. Its equivalence to 
the extreme BTZ black hole \cite{btz} can be shown \cite{pope, real}. To 
see 
this, first scale the $v$ and the $t$ coordinates such that the constants
$f_0$ and $h_0$ are set to one, and next define $H=\r$ as the new 
radial coordinate in (\ref{firstmetric}). Then we get
\be
ds^2= -(4m^2\r^2 - 2mh_1)dt^2 + h_1 d \b dt + \r^2d\b^2 + \frac{4\r 
^2}{(4m\r^2 -h_1)^2} d\r^2  \, , \hs{5} \b \equiv v- pt \, ,
\la{btzmetric}
\ee
where we have also performed the $t \rightarrow (\sqrt{m}/p) t$, 
$\r \rightarrow 2 \sqrt{m} \r$ and $\b \rightarrow [1/ (2 \sqrt{m})] \b$
scalings. Now (\ref{btzmetric}) is the extreme BTZ metric with  
total angular momentum $J=h_1$ and the total mass $M=2mJ=2mh_1$.

When $h_2 \neq 0$, this spacetime is another pp-wave in $AdS_3$. 
It is clear that it exists only for a non-zero scalar field. In this case, 
if we perform the coordinate change $H = \r$ in (\ref{firstmetric}), 
then when the $h_2$ term is negligible, the extreme BTZ metric 
(\ref{btzmetric}) receives corrections involving $(\ln \r)$ terms which
prevents the existence of a horizon. This signifies that this solution 
might be interpreted as a self-gravitating soliton \cite{it}.

Another complete solution for (\ref{a3}) and (\ref{a4}) is found when 
\be
\frac{H'}{F} =h_0 \, ,
\ee
where $h_0$ is a constant. Defining a new radial coordinate $H=\r$, 
one then finds supersymmetric solutions that were already studied in 
\cite{it}. In this case, the vector field components turn out to be 
constants. 
Unfortunately, we couldn't solve (\ref{a3}) and (\ref{a4}) in their 
full generality. 

\section{Conclusions}

We would now like to make some remarks about our results and suggest 
some future directions. In this paper, we have found new supersymmetric
solutions in addition to the already known ones. For the non-compact 
sigma model, we gave an explicit solution which we interpret as a 
non-topological soliton. To our knowledge, this is the first example 
of such an exact solution for a nonlinear sigma model coupled to 
supergravity. In addition, we also found a locally flat solution for the 
compact sigma model and a solution that we termed `string-like' for both 
types of sigma models. As we discussed above, these `string-like' solutions 
are obtained when $w_0=0$ and they resemble the string solution 
(without waves) given in \cite{string}. The differences lie in an extra 
term in the metric which doesn't affect the curvature invariants and 
the form of the vector field. This suggests a relationship between these 
two solutions which would be interesting to understand. 

We observed that in the flat sigma model limit ($a^2 \ra 0$) of our 
theory, the BTZ \cite{btz} black hole solution arises \cite{it}. However 
for $a^2 \neq 0$, we weren't able to find a solution which we could label 
as a black hole. This type of solution may not be allowed when the 
non-linear sigma model scalars are active. It is desirable to see 
whether this is true, by finding out all possible supersymmetric 
solutions as was done recently in some higher dimensional models 
(for a review see \cite{gaunt}.) One further generalization that is 
worth studying is to allow the coupling of more than one matter 
multiplets to supergravity.

In all our solutions $AdS$ space emerged when we took certain limits, 
which makes these useful for studying the $AdS/CFT$ duality. 
Half supersymmetric, smooth solutions recently attracted much attention 
after the appearance of \cite{llm} where it was shown that such solutions 
correspond to droplets in phase space occupied by the fermions on the $CFT$ 
side. Unfortunately, the $CFT$ dual of the model that we studied is not known 
yet. Once that is established, our explicit soliton solution might be 
important from this perspective.

We hope that our results will be useful in studying supersymmetric 
solutions of more complicated supergravity theories which are 
coupled to sigma model systems in higher dimensions, as well as 
in $D=3$ (for a review see \cite{witnicsam}). It is especially 
attractive to study the $D=3$, $N=8$ model \cite{n8}, since its 
higher dimensional origin and its dual $CFT$ are well-known.

Another thing to consider would be the explicit calculation of the 
conserved charges for the non-topological soliton that we found. 
It would be interesting to investigate the energy bound \cite{it, gito}
for this soliton. We hope to report on these issues soon.

\section*{Acknowledgments}

We would like to thank Rahmi G{\"u}ven and Ergin Sezgin for useful 
discussions. This work is partially supported by the Scientific and 
Technological Research Council of Turkey (T{\"U}B{\.I}TAK). The work of 
N.S. Deger is partially supported by the Turkish Academy of Sciences 
via The Young Scientist Award Program (T{\"U}BA-GEBIP).

\appendix

\section{\label{appen} Analysis of the Supersymmetry Breaking Condition}

In this appendix, we give the technical details that lead to 
(\ref{eqn4})-(\ref{eqn6}). We note that our analysis of \( \d 
{\tilde \psi}_\m =0 
\) 
goes along the lines of \cite{it}.

By defining \( B_\m = \frac{1}{4} \o_{\m}\,^{ab} \c_{ab} + m C^2 \c_{\m} 
\),
one can write the Killing spinor equation (\ref{susy1}) as
\be 
D_\m \ve \equiv \left( \del_\m + B_\m - \frac{i}{2 a^2} Q_\m 
\right) \ve = 0 \, . \la{kill}
\ee
Now imposing \( [D_\m, D_\n] \ve = 0 \), it can be shown that
this integrability condition is equivalent to
\be
(S^{\s}\,_{c} \c^c + G^\s) \ve = 0 \, , 
\la{integ}
\ee
where we have defined
\[ S^{\s}\,_{c} \equiv \e^{\m\n\s}(\del_\m B_{\n c} 
+ \e_{abc} B_{\m}\,^{a} B_{\n}\,^{b} ) \quad \mbox{and} \quad
G^\s \equiv - \frac{i}{2a^2} \e^{\m\n\s}\del_\m Q_\n \]
for simplicity. For a 1/2 supersymmetric solution, we now assume a 
projection of the form
\be
(1-\c^a b_a)\ve= 0 \, , 
\la{proj}
\ee
where $b_a$'s are some \emph{complex} functions that satisfy
\( b^a b_a=1 \). 
This condition can easily be solved with a spinor of the form
\be
\ve = N(1+ \c^a b_a) \ve_0 \, , \la{spin}
\ee
where $N$ is an arbitrary complex function and $\ve _0$ is a constant 
spinor. Inserting this solution for the spinor into the integrability 
condition (\ref{integ}), one finds
\be
S^{\s}\,_{c} + \e _{abc} S^{\s a} b^b - S^{\s}\,_{a} b^a b_c = 0 \, .
\la{seqn}
\ee

One now calculates the components of $S^{\s a}$ and $B_{\m}\,^{a}$
using the ansatz (\ref{metric}) for the metric and substitutes these
to the condition (\ref{seqn}) which yields
\bea
\frac{B_{t2}}{B_{\th 2}} = \frac{B_{t0}}{B_{\th 0}} &=& \L(r) \quad 
\mbox{for some function} \; \L(r) \, , \nonumber \\
S^{t2} (b_1 - b_2 b_0) - S^{t0} (1+(b_0)^2) &=& 0 \, , \la{a1} \\
S^{\th 2} (b_1 - b_2 b_0) - S^{\th 0} (1+(b_0)^2) &=& 0 \, . \nonumber
\eea
On the other hand, going back to the original Killing spinor equation
(\ref{kill}) and using (\ref{spin}) leads to
\bea
\del_\m (\ln{N}) - \frac{i}{2 a^2} Q_\m + B_{\m c} b^c &=& 0 \, , 
\la{beqn} \\
\del_\m b_a + B_{\m a} - B_{\m c} b^c b_a - \e_{abc} b^b B_{\m}\,^{c} &=&
0 \, . \la{ceqn}
\eea
However, (\ref{a1}) together with the $\m=t$ and $\m=\th$ components 
of (\ref{ceqn}) imply that
\be
(\del_t - \L(r) \del_\th) b_a = 0 \, . \la{deqn}
\ee

Now taking the $r$ derivative of (\ref{deqn}) and using the information
inherent in the $\m=r$ component of (\ref{ceqn}) implies that 
\( \del_\th b_a = 0 \), and thus \( \del_\t b_a = 0 \), unless
\( \L'(r) = 0 \). In what follows we assume that \( b_a = b_a(r) \)
only and keep $\L(r)$ arbitrary so that one is led to
\[ b_0 = b_2 Z \pm \sqrt{Z^2-1} \; , \quad b_1 = -Z \mp b_2 \sqrt{Z^2-1} 
\,, \]
where
\be
Z(r) \equiv \frac{B_{\th 2}}{B_{\th 0}} = - \frac{1}{2} \frac{H^2 G'}{F 
H'} +
2 m C^2 \frac{H}{H'} \la{ieqn}
\ee
for our choice of metric functions (\ref{metric}). Using these with 
the $\m=r$ component of (\ref{ceqn}), one also finds that
\[ b_2^{\prime} = \pm B_{r1}(r) (1-(b_2)^2) \sqrt{Z^2-1} \quad
\mbox{and} \quad
b_1^{\prime} = B_{r1}(r) ((b_1)^2 -1) \, , \]
which lead to
\be
Z' + 2 B_{r1}(r) (Z^2-1) = 0 \, . \la{feqn}
\ee 
Now using the explicit components of $B_{\m a}$ in (\ref{a1}), one also
obtains
\be
B_{r1}(r) = \frac{F'}{2 F Z} \, , \la{heqn}
\ee
which finally yields (\ref{defnz}) thanks to (\ref{feqn}).

For the metric ansatz employed, the explicit form of $B_{r1}(r)$ is
\[ B_{r1}(r) = \frac{1}{4} G' \frac{H}{F} + m C^2 \, . \]
Using this in conjunction with (\ref{heqn}) and (\ref{ieqn}) leads to
(\ref{eqn4}) and (\ref{eqn5}), together with
\[ \L(r) = g_0 + 4 m C^2 \frac{F}{H'} \quad \mbox{where} \; g_0 \;
\mbox{is an integration constant,} \]
after a careful scrutiny. All of these can be used in 
the $\m=r$ component of (\ref{ceqn}) to obtain the complex functions
$b_a$ as
\bea
b_2 &=& \frac{\b \left( \frac{2}{F} (-ik + \sqrt{F^2-k^2})\right)^p -1}
{\b \left( \frac{2}{F} (-ik + \sqrt{F^2-k^2})\right)^p +1} \, , \nonumber 
\\
b_0 &=& \pm i \frac{k}{F} + \frac{p}{F} b_2 \sqrt{F^2-k^2} \, ,
\quad \mbox{where} \; p^2=1 \,, \la{baeqn} \\
b_1 &=& - \frac{p}{F} \sqrt{F^2-k^2} \mp i \frac{k}{F} b_2 \, , \nonumber
\eea
for some integration constant $\b$.

Going back to (\ref{beqn}), one can now determine the function $N$ in
(\ref{spin}) as
\[ \ln{N} = \frac{1}{2} \ln{F} \pm i p \frac{k}{2} I(F(r)) 
+ \tilde{n}(\th,t) \, ,\]
where $\tilde{n}(\th,t)$ is an arbitrary function to be determined and
\be
I(F(r)) = \left\{
\begin{array}{lr}
\frac{1}{2k} \left\{2 \arctan{\left[ \frac{4 \b k}{(4 \b^2-1)F} \right] } 
+ i \left( \ln{\left[ \frac{F^2}{(1-4\b^2)^2 F^2+16\b^2 k^2} \right]} 
\right. \right. \\ \left. \left. \quad - 2 
\ln{\left[ \frac{2(4\b^2-1)[4\b F +ik+\sqrt{F^2-k^2}+
4\b^2(-ik+\sqrt{F^2-k^2})]}{(1+4\b^2)^2 ((4\b^2-1)F+4 i \b k)} \right]} 
\right) \right\} \, , & p=+1 \\
\frac{1}{2k} \left\{2 \arctan{\left[ \frac{4 \b k}{(\b^2-4)F} \right] }
- i \left( \ln{\left[ \frac{F^2}{(\b^2-4)^2 F^2+16\b^2k^2} \right]} 
\right. \right. \\ \left. \left. \quad + 2
\ln{\left[ \frac{2(\b^2-4)[-4\b F +\b^2 (-ik+\sqrt{F^2-k^2})+
4(ik+\sqrt{F^2-k^2})]}{(\b^2+4)^2 ((\b^2-4)F-4 i \b k)} \right]} 
\right) \right\} \, , & p=-1
\end{array} \right. \,. \la{ifeqn}
\ee
However, using the $\m=t$ and $\m=\th$ components of (\ref{beqn}),
one can show that $\tilde{n}(\th,t)$ necessarily has the form
\( \tilde{n}(\th,t) = i (c_1 \th + c_2 t) \) for some constants
$c_1$ and $c_2$.

All of the steps taken so far has been verified to be consistent
in themselves. Putting things together, the Killing spinor is 
finally obtained as
\be
\ve = \sqrt{F(r)} e^{i (c_1 \th + c_2 t)} e^{i pqk I(F(r))/2} (1+ \c^a 
b_a) 
\ve_0 \, ,
\ee
where $b_a$'s and the function $I(F(r))$ are given in (\ref{baeqn})
and (\ref{ifeqn}), respectively. Furthermore, the constants $c_2$ and
$c_1$ can also be used in finding the components of the vector field
$A_\m$ (\ref{ans2}), which yield (\ref{eqn1}) and (\ref{eqn2}), 
respectively.

One is now left with the other supersymmetry transformation 
$\d {\tilde \l} = 0$ (\ref{susy2}). It can easily be verified that this 
can 
be written in the form
\[ (\X^a \c_a + \X) \ve = 0 \, , \]
where 
\[ \X^0 = \frac{i}{F} ((\l-\e \psi)-G(n-\e \chi)) \, , \;\;
\X^1 = \frac{R'}{R} \, , \;\; \X^2 = \frac{i}{H} (n-\e \chi) \quad
\mbox{and} \;\; \X = 4 \e m a^2 C \, . \]
However, one easily finds that the projection condition used
for the Killing spinor (\ref{proj}) implies that \( \X^a b_a + \X = 0 \)
as well. Using \( b^a b_a=1 \) and the standard $\c_a$ algebra, one
also finds that $\X^a$ are related by
\[ \X b_a + \X_a + \e_{abc} \X^b b^c = 0 \, . \]
A careful study of these equations finally shows that there are in fact
only two conditions that are imposed by (\ref{susy2}) on the metric and
matter field components, which are simply (\ref{eqn3}) and (\ref{eqn6}).

This concludes our analysis for the simultaneous vanishing of the 
supersymmetry transformations (\ref{susy1}) and (\ref{susy2}) for our 
ansatz 
(\ref{metric}), (\ref{ans1}) and (\ref{ans2}), and the
derivation of the equations (\ref{eqn4}) through (\ref{eqn6}).

\end{document}